\documentclass[twocolumn,aps,prl,showpacs,byrevtex]{revtex4-1} %superscriptaddress,
\usepackage{epsf,graphicx,amssymb}
\usepackage[usenames]{color}
\usepackage{natbib}
\usepackage{float,ulem}
\usepackage{upgreek}
\usepackage{gensymb}
\usepackage{amsmath,amssymb}
\usepackage{color}
\usepackage{hyperref}
\hypersetup{
colorlinks=true, %colorise les liens
citecolor=magenta,
breaklinks=true, %permet le retour a la ligne dans les liens trop longs
urlcolor= blue, %couleur des hyperliens
linkcolor= blue, %couleur des liens internes
bookmarksopen=true, %si les signets Acrobat sont crees,
}

\definecolor{red}{rgb}{0.85,0.14,0.05}
\definecolor{green}{rgb}{0.23,0.62,0.14}
\definecolor{blue}{rgb}{0.05,0.35,0.85}

\begin{document}

\title{Freely Decaying Saffman Turbulence Experimentally Generated  by Magnetic Stirrers}
\author{Jean-Baptiste Gorce}
\email[E-mail: ]{jean-baptiste.gorce@u-paris.fr}
\author{Eric Falcon}
\affiliation{Universit\'e Paris Cit\'e, CNRS, MSC Laboratory, UMR 7057, F-75 013 Paris, France}

\date{\today}
\begin{abstract}
We investigate experimentally the decay of three-dimensional hydrodynamic turbulence, initially generated by the erratic motions of centimeter-size magnetic stirrers in a closed container. Such zero-mean-flow homogeneous isotropic turbulence is well suited to test Saffman’s model and Batchelor’s model of freely decaying turbulence. Here, we report a consistent set of experimental measurements (temporal decay of the turbulent kinetic energy, of the energy dissipation rate, and growth of the integral scale) strongly supporting the Saffman model. We also measure the conservation of the Saffman invariant at early times of the decay and show that the energy spectrum scales as $k^2$ at large scales and keeps its self-similar shape during the decay. This letter  thus presents the first experimental evidence of the validity of the connection between the Saffman invariant and the $k^2$-energy spectrum of the large scales. The final decay regime closely corresponds to Saffman’s model when the container size is sufficiently large.
\end{abstract} 
%by using 1-cm magnetic stirrers that initially inject the energy within the volume of a closed container to generate zero-mean-flow homogeneous isotropic turbulence.

\maketitle

\paragraph*{Introduction.\textemdash}The decay of three-dimensional (3D) turbulent flows has been extensively investigated to comprehend the energy transfer and the dynamics of the large scales, the scales larger than the forcing scale~\cite{Davidson}. Understanding the decay rate of turbulent kinetic energy is important for fundamental theories, numerical simulations of turbulence and applications such as weather forecasting or energy harvesting. However, the physical mechanisms that control the decay rate of fully developed homogeneous turbulence are not clearly identified~\cite{Davidson}.

Currently, the Batchelor model~\cite{Kolmogorov,Batchelor56} and Saffman model~\cite{Saffman67} have competing hypotheses to describe the decay of homogeneous turbulence. Both models assume distinct invariants depending on how the turbulence is initially generated, and this distinction is reflected in the scaling of the energy spectrum at large scales. Specifically, a turbulent flow with significant linear momentum possesses an energy spectrum at large scales given by $E(k)\sim k^2$ (Saffman)~\cite{Saffman67}. Conversely, a turbulent flow initially generated by a strong angular impulse and a negligible linear impulse exhibits a $E(k) \sim k^4$ energy spectrum at large scales (Batchelor)~\cite{Batchelor56}.

Both types of turbulence can be generated in direct numerical simulations ~\cite{Davidson,Lesieur,IshidaJFM2006,DavidsonJFM2012,YoshimatsuPRF2019,Anaspof2020}, and this raises questions about how the initial conditions or energy injection methods control the decay of turbulent flows. Direct numerical simulations  studies on freely decaying turbulence impose the spectrum at large scales using a Gaussian process to inject energy~\cite{Kaneda04}, while the small scales are not turbulent and do not exhibit a $k^{-5/3}$ power-law spectrum.

Experimental open systems, such as grid turbulence, can  reach a Reynolds number up to $5 \times 10^6$~\cite{Sinhuber2015} and are plausible candidates to measure the decay of turbulence. However, they possess a mean flow and different decay rates were then reported using passive grids~\cite{Batchelor56,Comte-Bellot66,Mohamed90,Krogstad2010,Sinhuber2015}, fractal grids with multiscale grids~\cite{Hurst2007,DavidsonJFM2011,Valente2011,Valente2012} or active grids~\cite{Burattini2005,Mazellier2008,Thormann2014}. On the other hand,  there exist complementary laboratory experiments in closed systems (where fans, loudspeakers, jets, or rotating elements energize the fluid) generating zero-mean-flow homogeneous isotropic turbulence (HIT) to study the decay of turbulence~\cite{Nezami2023}. However, the decay rate in such closed systems is influenced by the different degrees of isotropy, the asymmetry of the forcing, or secondary large-scale flows~\cite{Nezami2023}. Indeed, the influence of a mean flow or secondary flows affects the energy budget and the time dependence of the turbulent kinetic energy, which stresses why isotropy is crucial to test the decay law~\cite{Moisy2011}. More direct evidence in zero-mean-flow HIT experimental setups is thus required to confirm Saffman's or Batchelor's model and to clarify the connection between the large-scale energy spectrum and the invariants of freely decaying turbulence.
%More direct evidence in experimental setups with isotropic and zero-mean flow is thus required to confirm the Saffman or Batchelor model.
%On the other hand, zero-mean flow grid-generated turbulence experiments have recently been performed in a rotating container~\cite{Morize2005}. It has been observed that as the angular velocity increases, the anisotropy of the velocity field impacts the energy decay rate as a function of the time~\cite{Moisy2011}. More direct evidence in experimental setups with isotropic and zero-mean flow is thus required to confirm the Saffman or Batchelor model.
%On the other hand,  grid-generated turbulence in closed experimental systems} have  been performed in a rotating container~\cite{Morize2005}. %It has been observed that as the angular velocity increases, the anisotropy of the velocity field impacts the energy decay rate as a function of the time~\cite{Moisy2011}. 

Here, we initially generate 3D hydrodynamic turbulence using centimeter-size magnetic stirrers immersed in a large liquid reservoir and we then halt the forcing to study freely decaying turbulence. The advantage of such volume forcing is to generate sufficient zero-mean-flow HIT required to compare Saffman's model and Batchelor's model of freely decaying turbulence. Using this technique, we report a consistent set of experimental observations (kinetic energy, dissipation rate, and integral scale) robustly supporting the Saffman model. We also measure the conservation of the Saffman invariant at early times of the decay. The energy spectrum scales as $k^2$ at large scales and conserves a self-similar shape during the decay.

\paragraph*{Theoretical backgrounds.\textemdash}Assuming that the energy spectrum $E(k,t)$ is analytic at $k = 0$, a Taylor expansion at small $kr$ (large scales) shows the following leading terms~\cite{DavidsonJFM2011}
\begin{align}
    E(k,t)=\frac{Lk^2}{4\pi^2}+\frac{Ik^4}{24\pi^2}+...
\label{spectrum}
\end{align}

\noindent with $L=\int_0^\infty \left< \mathbf{u}\left(\mathbf{x},t\right)\cdot\mathbf{u}\left(\mathbf{x+r},t\right)\right>dr $ is Saffman’s integral, a measure of the linear momentum held in the turbulence~\cite{Davidson2011}, $I=-\int_0^\infty \left< \mathbf{u}\left(\mathbf{x},t\right)\cdot\mathbf{u}\left(\mathbf{x+r},t\right)\right>r^2dr$ is Loitsyansky's integral, suggested to be related to the angular momentum~\cite{Landau} and $\left< \mathbf{u}\left(\mathbf{x},t\right)\cdot\mathbf{u}\left(\mathbf{x+r},t\right)\right>$ the autocorrelation function of the velocity field $\mathbf{u}$~\cite{Davidson,Davidson2011}. In fully developed freely decaying HIT, $L\sim u^2l^3$ with $l$ the integral scale defined as $l=\int_0^\infty f(r,t)dr$, where $f(r,t)$ is the longitudinal velocity autocorrelation function. Unlike $L$, the integral $I$ is not, in general, an invariant during the initial decay~\cite{Davidson,Batchelor56,Proudman}.

The decay rate of the squared velocity fluctuations $u^2=\left<\mathbf{u}^2\right>/3$ can be evaluated by assuming that $du^2/dt$ is equal to minus the dissipation rate $-\epsilon$~\cite{Kolmogorov}
\begin{equation}
\frac{du^2}{dt}=-\epsilon=-C \frac{u^3}{l}
\label{eq2}
\end{equation}

\noindent with $C$ a constant of order unity, which depends on the Taylor Reynolds number and the large-scale forcing procedures~\cite{Sreenivasan,Lohse,Sreenivasan1998,Kaneda2003,Vassilicos2015}. Using the invariant $u^2l^3$ (Saffman) or $u^2l^5$ (Batchelor), the time dependence of $u^2$, $l$, and $\epsilon$ can be derived~\cite{Davidson,Davidson2011}, as summarized in Table~\ref{table1}. The decay of the kinetic energy during the final period of decay is also shown in Table~\ref{table1}.

\begin{table}[h!]
\begin{ruledtabular}
\begin{tabular}{ccc}
Model & Saffman & Batchelor \\
Large-scale spectrum  & $E(k)\sim k^2$ & $E(k)\sim k^4$ \\
\hline
\textit{Initial decay} \\
Invariant & $L\sim u^2l^3$ & $I\sim u^2l^5$ \\
\\
$u^2/u_0^2$ & $\left(1+at\right)^{-6/5}$ & $\left(1+bt\right)^{-10/7}$\\
$l/l_0$ & $\left(1+at\right)^{2/5}$ & $\left(1+bt\right)^{2/7}$\\
$\epsilon/\epsilon_0$ & $\left(1+at\right)^{-11/5}$ & $\left(1+bt\right)^{-17/7}$\\
\hline
\textit{Final decay} \\
$u^2$ & $\left(t-t_*\right)^{-3/2}$ & $\left(t-t_*\right)^{-5/2}$\\
\end{tabular}
\end{ruledtabular}
\caption{Time evolution of $u^2$, $l$, and $\epsilon$ during the initial decay and of $u^2$ during the final decay depending on the initial conditions of the turbulent flow. The large-scale energy spectrum $E(k)\sim k^2$ corresponds to Saffman's model and $E(k)\sim k^4$ corresponds to Batchelor's model. The values of the constants are $a=\frac{5}{6} C \frac{u_0}{l_0}$ and $b=\frac{7}{10} C \frac{u_0}{l_0}$. The initial values are indexed with 0: $u_0$, $l_0$ and $\epsilon_0$.}
\label{table1}
\end{table}

\paragraph*{Experimental setup.\textemdash}Experiments are carried out in two different fluid square containers sealed by a transparent lid. The dimensions are $ 11 \times 11  \times 8$ cm\textsuperscript{3} (small tank) and $ 33 \times 33  \times 20$ cm\textsuperscript{3} (large tank) (see the schematics in the Supplemental Material~\cite{supmatt}). The choice of these varying sizes allows for assessing finite-size effects in the experimental observations. In the small tank, measurements are taken using two different liquids: either water or a lower-viscosity liquid (Novec) while exclusively water is used for measurements in the large tank. Both fluids are seeded with hollow glass sphere fluid tracers (10 $\upmu$m, concentration of 0.21 ppm) illuminated by a horizontal laser sheet, and a high-speed camera (Phantom v1840) records high-resolution movies ($2048 \times 1952$ pixels\textsuperscript{2}) at a range of speeds 100–400 fps. Energy is transferred into the fluid  by the continuous erratic motions of $N$ magnetic stirrers (1 cm in size) driven by a monochromatic vertical magnetic field of frequency $F$~\cite{Falcon2013,Falcon2017,Gorce2023}, which generate a turbulent flow \cite{Cazaubiel2021,Gorce2022}.

The control parameters in the small tank are the number of magnetic stirrers $N=50$, the frequency of the oscillating magnetic field $F=50$ Hz, and the magnetic field intensity $B=240$ G and correspond to the maximal values of this system (see the illustrative movie in the Supplemental Material~\cite{supmatt}). The typical rms velocity of the magnetic stirrers in water is 20 cm/s~\cite{Gorce2023}. The control parameters in the large tank are $N=450$, $F=20$ Hz, and $B=360$~G.

At $t=0$, turning off the magnetic field stops the energy injection and settles the stirrers at the container's bottom. During this transient regime of turbulent decay, a nonintrusive particle image velocimetry (PIV) technique \cite{adrian1991particle} using the PIVlab algorithm \cite{Thielicke2014} measures the fluid velocity field in the $xy$ horizontal plane. For the small tank, the initial value of the standard deviation of the fluid velocity is equal to $u_0=6.6$ cm/s, giving the initial Reynolds number $\mathrm{Re_0}=u_0l_0/\nu_w=3000$, with $l_0=5$ cm the initial integral length scale and $\nu_w=10^{-6}$ m\textsuperscript{2}/s the kinematic viscosity of water.

\begin{figure}[!t]
\begin{center}
\includegraphics[width=8.6cm]{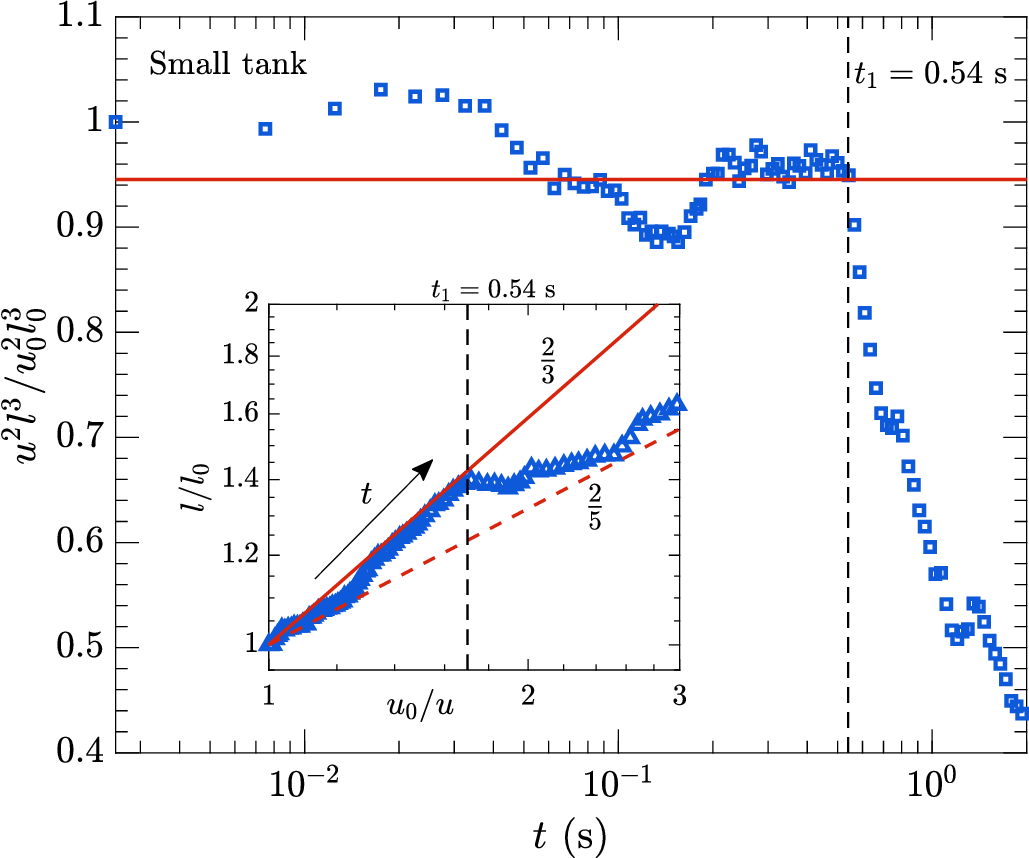} 
\caption{Time evolution of Saffman invariant $u^2 l^3$ using water as working fluid. The solid line represents the mean value of the invariant up to $t_1=0.54$ s. Inset: $l/l_0$ as a function of $u_0/u$. The equation of the solid line is $y=\left(u_0/u\right)^{2/3}$ (Saffman) and the dashed line is $y=\left(u_0/u\right)^{2/5}$ (Batchelor). The black arrow represents the direction of time and the dashed line gives the time $t_1$ after which $u^2l^3$ decreases significantly.}
\label{invariant}
\end{center}
\end{figure}

\paragraph*{Mean-flow free, homogeneity, and isotropy.\textemdash}Using the horizontal velocity fluctuations $u_x$ and $u_y$, the structure functions $S_2^{u_x}(r)=\langle[u_x(x+r)-u_x(x)]^2\rangle_x$ and $S_2^{u_y} (r)$ are measured nearly identical, illustrating the homogeneity and isotropy of the velocity field during the decay in the small tank (see Supplemental Material~\cite{supmatt}). The isotropy coefficient is also measured using the ratio of the standard deviations.$\sigma_{u_x}/\sigma_{u_y}$ is equal to $1\pm0.004$ on average during the decay. The ratio of the mean velocity and standard deviation, $\left<u_x\right>/\sigma_{u_x}$ and $\left<u_y\right>/\sigma_{u_y}$, are 2.2\% and 7\%, respectively (see Supplemental Material~\cite{supmatt}), confirming the isotropy, the absence of mean and secondary flows.

\begin{figure}[!t]
\begin{center}
\includegraphics[width=8.6cm]{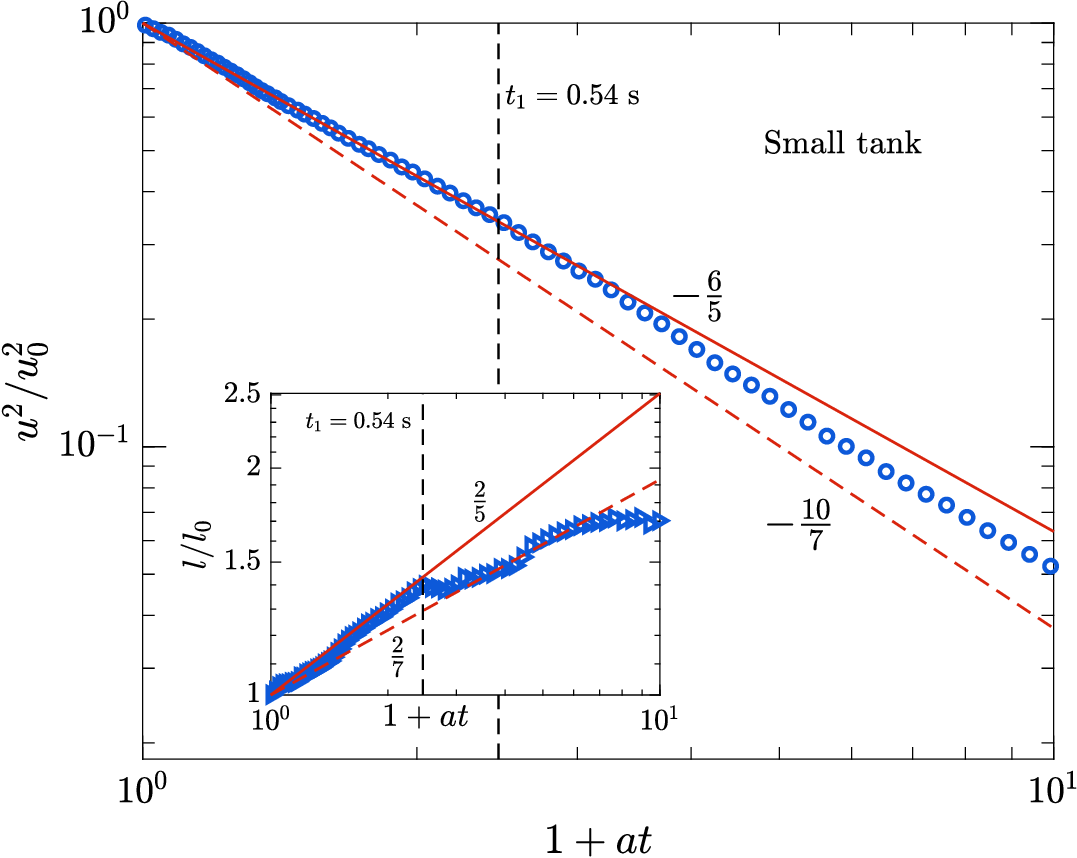} 
\caption{Decay of the squared velocity fluctuations $u^2$ as a function of the rescaled time $1+at$ (water). The solid line corresponds to a power law defined as $\left(1+at\right)^{-6/5}$ (Saffman) and the dashed line represents the power law $\left(1+at\right)^{-10/7}$ (Batchelor). Inset: time evolution of the integral scale $l$. The solid line represents the power law $\left(1+at\right)^{2/5}$ (Saffman) and the dashed line $\left(1+at\right)^{2/7}$ (Batchelor).}
\label{integral}
\end{center}
\end{figure}

\paragraph*{Initial decay.\textemdash}The initial decay in the small tank is evaluated using exclusively water.PIV measurements of the horizontal $u_x$ and vertical velocity $u_z$ between $z=6$ to $8$~cm confirm the turbulent decay is not affected by the downward motion of the stirrers (see Supplemental Material~\cite{supmatt}). Figure~\ref{invariant} shows that the quantity $u^2l^3$ is invariant at the beginning of the decay until it decreases after a time $t_1=0.54$ s. This illustrates the invariance of Saffman's integral $L\sim u^2 l^3$ and the conservation of linear momentum during the initial decay ($t<t_1$). The present measurement supports the hypothesis that the magnetic stirrers inject strong linear momentum into the turbulent eddies ($L>0$), which is also endorsed by the comparison of the time evolutions of the quantities $u^2l^3$ and $u^2l^5$ shown in Supp. Mat.~\cite{supmatt}. The inset of Fig.~\ref{invariant} also illustrates a power-law relationship between $1/u$ and $l$ with a 2/3 slope consistent with Saffman’s theory, as indicated by the solid line.

\begin{figure}[!t]
\begin{center}
\includegraphics[width=8.6cm]{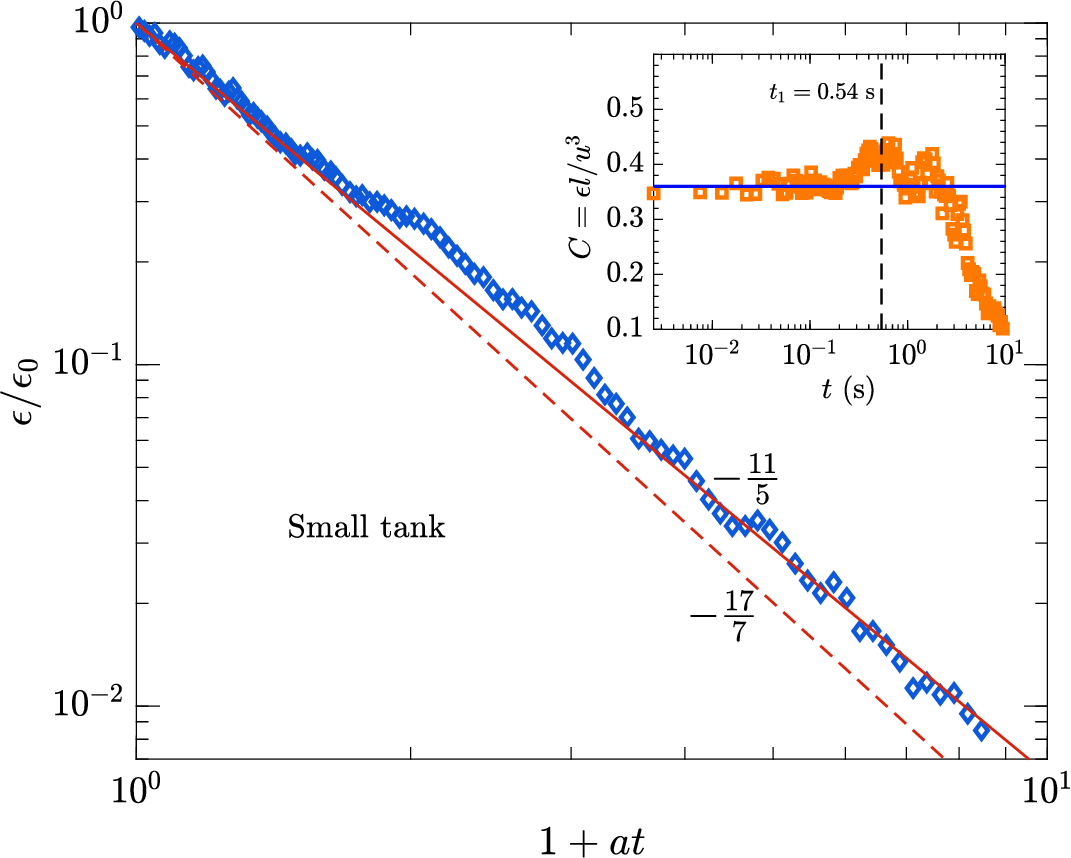} 
\caption{Decay of the energy dissipation rate $\epsilon$ as a function of the rescaled time $1+at$, with water as working fluid. The solid line represents $\left(1+at\right)^{-11/5}$ (Saffman) and the dashed line $\left(1+at\right)^{-17/7}$ (Batchelor). The initial value of the dissipation rate is $\epsilon_0=2.1 \times 10^{-3}$ m\textsuperscript{2}/s\textsuperscript{3}. Inset: time evolution of the constant $C$ measured from the ratio $\epsilon l/u^3$. The solid line represents the mean value of $C$ for $t\leq t_1$.}
\label{dissipation}
\end{center}
\end{figure}

The measurements shown in Fig.~\ref{invariant} suggest a potential Saffman turbulence scenario (second column in Table~\ref{table1}) in which the turbulent kinetic energy should decay as $u^2/u_0^2 = \left(1+at\right)^{-6/5}$ and the integral length scale increases as $l/l_0= \left(1+at\right)^{2/5}$, with $a=5C u_0/(6l_0)$. The value $a=2.7$ s\textsuperscript{-1} is inferred from the initial values of $u_0$ and $l_0$, and the constant   $C=0.37\pm 0.02$ measured from Eq.~\eqref{eq2}. %$C$ is of the order of 1 as in most turbulent flows~\cite{Sreenivasan1998}.} 
A correct definition of this value is essential for accurately assessing the time dependence of $u$ and $l$ during the decay~\cite{Mohamed90}.  %and is influenced by large-scale forcings~\cite{ Kaneda2003}. %$C$ exhibits a small variation of around 6\%, such that we use the mean value of $C$ over the 0.54 s time interval.

Figure~\ref{integral} shows the decay of $u^2/u_0^2$ as a function of the rescaled time $1+at$. It confirms the power-law relationship between these two quantities and the agreement with Saffman’s model for $t \leq t_1$. The inset of Fig.~\ref{integral} illustrates that the integral length scale $l$ increases during the decay and then saturates at $\left(1+at\right) \approx 6$ (i.e., $t \approx 1.85$~s). For $t \leq t_1$, $l/l_0$ is well fitted by the solid line given by Saffman's model and depicts a stronger increase in $l$ than in Batchelor's model. Deviations of $u^2$ and $l$ from the Saffman laws (solid lines) are observed after a time $1+at_1=2.4$ because the size of the biggest eddies [$l(t_1)=7$~cm] becomes comparable with the size of the container.

The rate at which the kinetic energy is dissipated is computed from the expression $\epsilon=15 \nu \langle \left( \partial u_x/ \partial x \right)^2 \rangle_{x,y}$, which is derived assuming HIT~\cite{Wang2021}. The measured initial dissipation rate is equal to $\epsilon_0=2.1 \times 10^{-3}$~m\textsuperscript{2}/s\textsuperscript{3}. Figure~\ref{dissipation} shows that the decrease of $\epsilon$ is in good agreement with Saffman's model. The measurements are very well fitted by $\left(1+at\right)^{-11/5}$, which is represented by the solid line in Fig.~\ref{dissipation}. The inset of Fig.~\ref{dissipation} represents the time evolution of the constant $C$ given by Eq.~\eqref{eq2}. This illustrates that $C$ is approximately constant up to $t=t_1$, suggesting that the velocity field is not fully turbulent after $t_1$ and that different physical mechanisms dissipate the turbulent kinetic energy of the liquid  such as dissipation at the tank boundaries.

\begin{figure}[!t]
\begin{center}
\includegraphics[width=8.6cm]{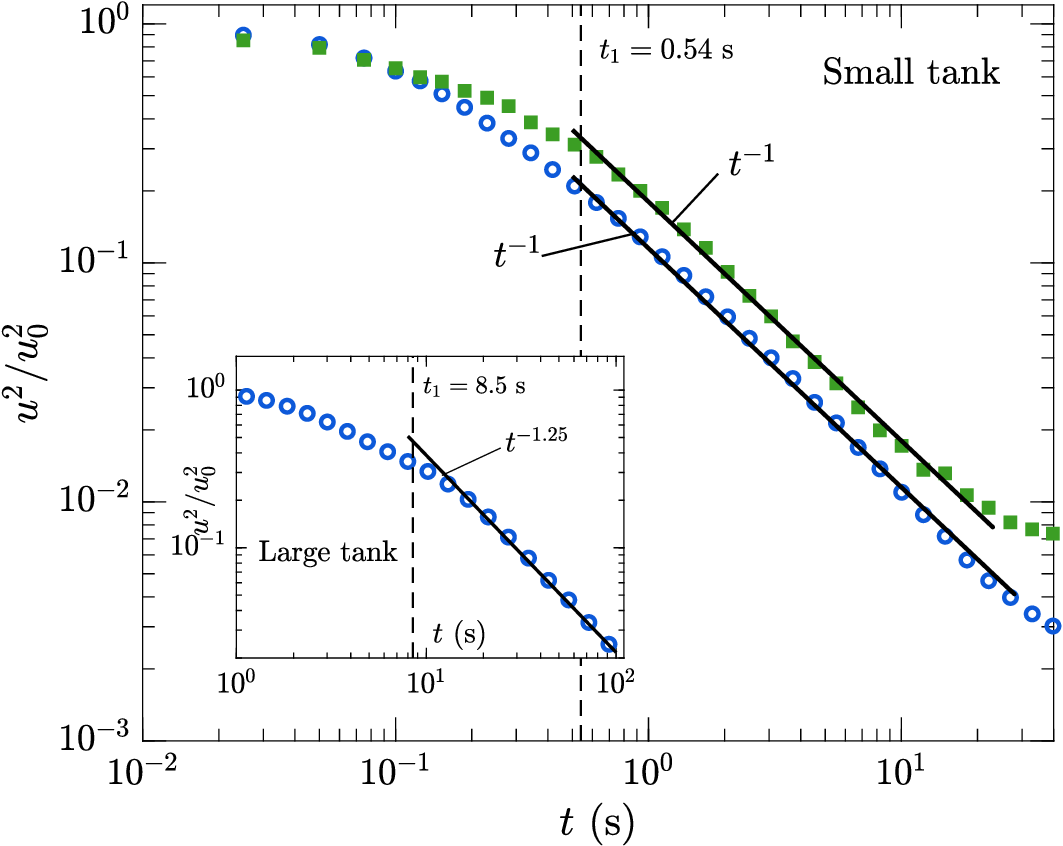} 
\caption{Decay of the turbulent kinetic energy in the small tank with two different fluids. The blue circles correspond to the measurement performed with water and the green squares correspond to Novec. The solid lines represent a $t^{-1}$ power law. Inset: measurements performed in the large reservoir filled with water. The solid line represents a $t^{-1.25}$ power law.} 
\label{kinetic}
\end{center}
\end{figure}

\paragraph*{Final decay.\textemdash}After $t_1$, the nonlinear inertial terms in the equations of motion are supposedly negligible and the dissipation of the turbulent kinetic energy solely depends on the viscosity $\nu$. The evolution of the turbulent kinetic energy during this final decay period can be derived from the initial large-scale spectrum (see Supplemental Material~\cite{supmatt}). As summarized in Table~\ref{table1}, the expression is given by either $u^2\sim\left(t-t_*\right)^{-3/2}$ for $E(k)\sim k^2$~\cite{Saffman67} or $u^2\sim\left(t-t_*\right)^{-5/2}$ for $E(k)\sim k^4$~\cite{Batchelor53}, where $t_*$ denotes some instant of time inside the final period~\cite{Batchelor53}. These power laws are derived under the assumptions that $\left(t-t_* \right) \rightarrow \infty$, which is challenging to achieve in experimental systems during the final decay stage. In addition, Ref.~\cite{Skrbek2000} pointed out that the value of the power-law exponent $\alpha$ in $\left(t-t_*\right)^{-\alpha}$ is highly sensitive to the choice of the virtual time parameter $t_*$. Consequently, we have chosen to directly fit the experimental data using a power-law model without introducing a virtual time origin $t_*$.

We conducted experiments in the small tank using two fluids (water or Novec) with different densities and viscosities to explore how these fluids dissipate turbulent kinetic energy during the final decay. The kinematic viscosity of Novec 7100 is $\nu_n=0.4 \times 10^{-6}$~m\textsuperscript{2}/s and its density is $\rho_n=1.5\times 10^{3}$~kg/m\textsuperscript{3}~\cite{novec}. Figure~\ref{kinetic} illustrates the decays of the turbulent kinetic energy with water (circles) and Novec (squares) that are both very well fitted by a $t^{-1}$ power law. The exponents of the power laws are independent of the kinematic viscosity $\nu$, which is consistent with the theoretical derivation (see Supp. Mat.~\cite{supmatt}). The deviation of the exponent from the value -3/2 derived in Saffman's model is likely due to the size of the biggest eddies [$l(t_1)=7$~cm] becoming comparable with the size of the container. This effect is known to alter the power-law exponent of the decay~\cite{Skrbek2000, Thornber2016, Meldi2017}. Additionally, finite Reynolds number effects contribute to this deviation~\cite{Anaspof2020}.

To reduce the finite-size effects of the small tank and the dissipation at its boundaries, we also conducted experiments within the large tank. The inset of Fig.~\ref{kinetic} shows that a $t^{-1.25}$ power law is observed for 1 order of magnitude. This supports the fact that the finite-size effects control the decay rate during the final period of decay and the time power-law exponent becomes closer to -3/2 (Saffman's model) in the large tank experiment. Note that the initial decay is not observed in the large tank because the initial Reynolds number is too small ($\mathrm{Re_0'}=u_0'l_0/\nu_w=650$, with $u_0'=1.3$~cm/s). Indeed, Fig.~\ref{spectrum} illustrates that the $k^{-5/3}$ power spectrum is no longer observed after only 0.01~s, which is clearly insufficient to resolve correctly the initial decay.

\begin{figure}[!t]
\begin{center}
\includegraphics[width=8.6cm]{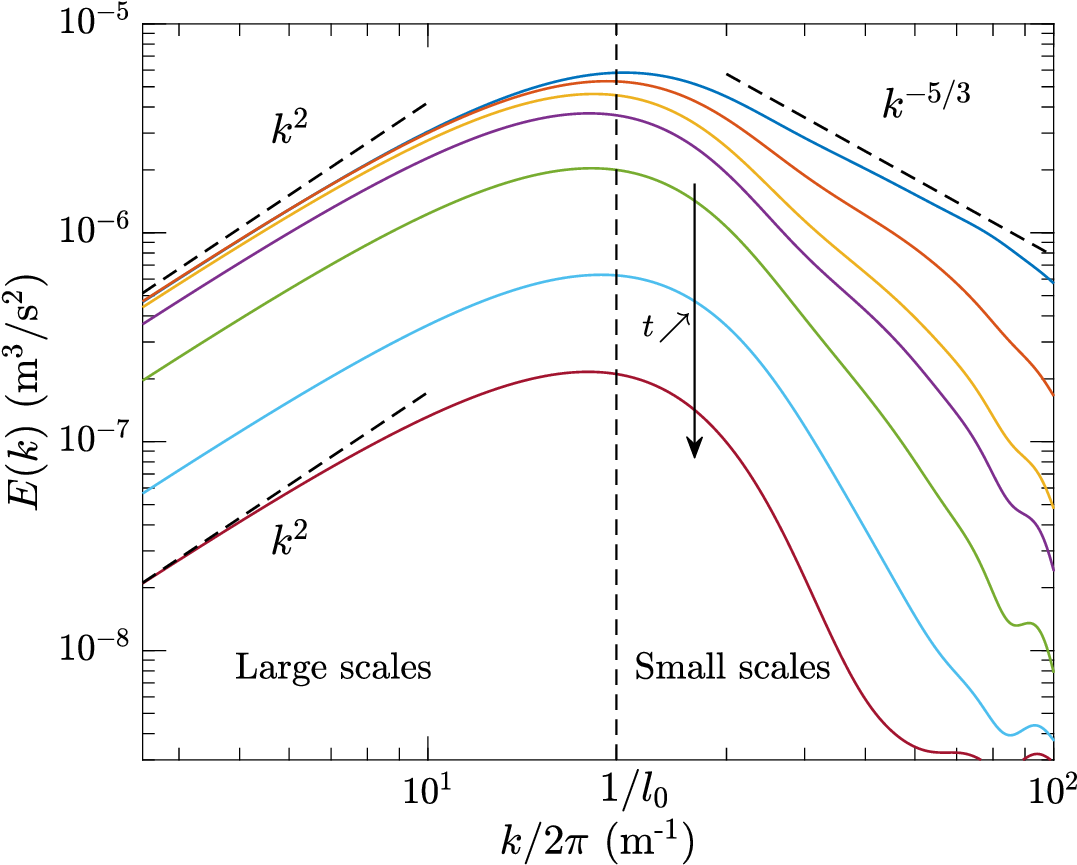} 
\caption{Decay of the energy spectrum in the large reservoir. The vertical dashed line corresponds to the initial inverse integral length $1/l_0$ separating the large and small scales. Here, $t=0,0.09,3.66,5.96,15.83,42.01$, and $104.99$ s.}
\label{spectrum}
\end{center}
\end{figure}

\paragraph*{Energy spectrum.\textemdash}In the absence of nonlinear transfer of energy across scales, Lin's equation, given by $\partial E(k,t)/\partial t \sim -2 \nu k^2 E(k,t)$, implies that the expected $k^2$ energy spectrum at large scales should persist over time throughout the decay. Measurements performed in the large tank confirm the conservation of the $k^2$ power law during the final decay stage, whereas the smaller scales lose their turbulent characteristics and exhibit a steeper power-law trend (Fig.~\ref{spectrum}). These observations align with the idea that viscosity dissipates the excess energy during the final decay and suggest that Saffman turbulence is observed here.

\paragraph*{Conclusion.\textemdash}We report on the freely decaying 3D turbulence, initially generated by the erratic motions of centimeter-size magnetic stirrers in a closed experimental setup. Such isotropic, mean-flow-free turbulence is well suited to compare Saffman and Batchelor models of freely decaying turbulence. Our experimental measurements (temporal decay of the turbulent energy kinetic, of the energy dissipation rate, and growth of the integral scale) robustly support Saffman model. Saffman invariant is also well conserved at early times of the decay. The energy spectrum scales as $k^2$ at large scales and conserves a self-similar shape during the decay.  This letter thus presents the first experimental evidence of the connection between Saffman invariant $L\sim u^2 l^3$ and the large-scale energy spectrum in $k^2$. The final decay is also reported in two different-size experimental systems. All these results support the existence of freely decaying Saffman turbulence involving turbulent eddies with a significant linear momentum input. Our results could be applied to physical, geophysical, or industrial turbulent flows with a finite mean flow and are of primary importance.

\begin{acknowledgments}
We thank A. Di Palma and Y. Le Goas, for technical help. This work was supported by the French National Research Agency (ANR LASCATURB project No. ANR-23-CE30-0043-02) and by the Simons Foundation MPS N$^{\rm o}$651463-Wave Turbulence.
\end{acknowledgments}

%%%%%%%%%%%%%%%%%%%%%%%%%%%%%%%%%%%%%%
%%%%%%%%%%%% REFERENCES %%%%%%%%%%%%%%%%%%
%%%%%%%%%%%%%%%%%%%%%%%%%%%%%%%%%%%%%%


\begin{thebibliography}{99}
\bibitem{Landau}$I$ is only a conserved quantity for a fast enough decay of $f(r)$ when $r \rightarrow 0$, and thus a direct consequence of the conservation of angular momentum. %L. Landau and E. Lifshitz, {\it Fluid Mechanics} (Elsevier Science, New York, 2013), Vol. {\bf6}.
\end{thebibliography}
\end{document}